\documentclass[journal]{IEEEtran}

\usepackage{graphicx, bm, mathrsfs}
\usepackage{cite,array, enumerate}
\usepackage{amssymb}
\usepackage{dsfont, bm, color}
\usepackage{epstopdf}
\usepackage[cmex10]{amsmath}
\usepackage{multirow}
\usepackage{booktabs}
\usepackage{threeparttable}
\newcommand{\overbar}[1]{\mkern 1.5mu\overline{\mkern-1.5mu#1\mkern-1.5mu}\mkern 1.5mu}
\usepackage{cases, stackengine}

\newtheorem{theorem}{Theorem}

\newtheorem{definition}{Definition}
\newtheorem{remark}{Remark}

\newcommand{\inte}{\textrm{Int}}

\begin{document}

\title{\LARGE \bf Multiple Control Functionals for Interconnected Time-Delay Systems
\thanks{This work was supported by the Fundamental Research Funds for the Central Universities under Grant DUT22RT(3)090, and the National Natural Science Foundation of China under Grant 61890920, Grant 61890921 and Grant 08120003.}
}

\author{Zhuo-Rui Pan, Wei Ren, and Xi-Ming Sun
\thanks{The first two authors contributed equally to this work.}
\thanks{Z.-R. Pan, W. Ren and X.-M. Sun are with the Key Laboratory of Intelligent Control and Optimization for Industrial Equipment of Ministry of Education, Dalian University of Technology, Dalian 116024, China. 
Email: \textrm{\small panzhuorui@mail.dlut.edu.cn, wei.ren@dlut.edu.cn, sunxm@dlut.edu.cn}}
}

\maketitle

\begin{abstract}
Safety is essential for autonomous systems, in particular for interconnected systems in which the interactions among subsystems are involved. Motivated by the recent interest in cyber-physical and interconnected autonomous systems, we address the safe stabilization problem of interconnected systems with time delays. We propose multiple control Lyapunov and barrier functionals for the stabilization and safety control problems, respectively. In order to investigate the safe stabilization control problem, the proposed multiple control functionals are combined together via two methods: the optimization-based method and the sliding mode based method. The resulting controllers can be of either explicit or implicit forms, both of which ensure the safe stabilization objective of the whole system. The derived results are illustrated via a reach-avoid problem of multi-robot systems. 
\end{abstract}\vspace{-10pt}

\section{Introduction}
\label{sec-introduction}

In the real world, complex engineering systems consist of a large number of subsystems with their interconnections and configurations \cite{Bonilla2020complexity}. The interconnection affects global behaviors which however may not be apparent from all subsystems \cite{Coogan2014dissipativity}, while the spatial configurations may result in time delays due to remote information transmission among all subsystems. Many physical systems like power systems and robotic systems \cite{Guiochet2017safety} can be modeled as interconnected time-delay systems, and how to address the effects of interconnection and time delays on system performances is still challenging. Among different system performances, safety and stabilization are fundamental and essential, which aims to guarantee all subsystems be stable and safe under some controllers. In particular, due to the interconnection among different subsystems and even the human-robot interaction \cite{Fang2023integrated, Guiochet2017safety, Ferraguti2022safety}, safety is of great importance for interconnected systems. Against the above background, in this paper we investigate the safety and stabilization problems of interconnected time-delay systems. 

Among different methods to deal with the safety and stabilization of dynamical systems, control Lyapunov and barrier functions have been extensively implemented in the literature \cite{Coogan2014dissipativity, Fridman2014tutorial, Ames2016control, Ferraguti2022safety}. With proper control Lyapunov and barrier functions, both centralized and distributed approaches have been proposed. However, the controller design in a centralized manner is inevitably limited to interconnected systems with moderate sizes, thereby resulting in the computational complexity. Furthermore, the controller design becomes more difficult if other issues like the interconnection and time delays are involved. Hence, the distributed approaches show a promising way for the controller design \cite{Wang2017safety, Panagou2015distributed}. Among different distributed approaches, the optimization-based methods are commonly-used due to their efficacy, and the key is to formulate the controller design into a distributed quadratic programming (QP) problem \cite{Ren2023vector, Wang2017safety}, which can be solved efficiently via many existing techniques. However, time-delay optimization problems may not be easy to be resolved and only numerical solutions could be derived. Other techniques, such as small-gain theorem \cite{Lyu2022small}, vector control functions \cite{Ren2023vector}, sliding mode control \cite{Ren2022Razumikhin2} and dissipativity analysis \cite{Coogan2014dissipativity}, are also effective to design the controllers explicitly.

Inspired by the above discussion, in this paper we investigate the safety and stabilization control problems of interconnected time-delay systems. In order to deal with the safety and stabilization control problems individually, multiple control Lyapunov and barrier functionals are proposed. In particular, each functional is assumed to consist of two parts. The one part involves the current state only, while the other part depends on the time-delay state trajectory. This setting is consistent with some existing works \cite{Ren2022Razumikhin2, Ren2022Razumikhin} and is to facilitate the distributed controller design. In addition, the coupling from the interconnection nature is embedded in the properties of multiple control functionals, which are orchestrated via the small-gain techniques. In this way, the distributed controller can be designed to achieve the stabilization objective of the overall system, whereas the safety objective is guaranteed via the existence of multiple control barrier functionals. Furthermore, we propose both implicit and explicit methods to address the safety and stabilization problems simultaneously. In the implicit method, an optimization problem is formulated via the combination of the derived distributed stabilizing controller and the quadratic programming. That is, the formulated optimization problem is solved to ensure the safety objective while tracking the derived distributed stabilizing controller. The explicit method is based on the construction of multiple sliding mode surface functionals, which further results in an explicit form of the distributed controller. In contrast to the interconnected case, our previous works \cite{Ren2022Razumikhin2, Ren2022Razumikhin} consider the centralized case only. In conclusion, in this paper we propose multiple control Lyapunov and barrier functions for interconnected time-delay systems and investigate the satisfaction of the safe stabilization objective, which extends existing control barrier functions to more general cases and provides novel ways for the safety control of interconnected systems. 

The problem is formulated in Section \ref{sec-nonconsys}. Multiple control functionals are proposed in Section \ref{sec-multiplefunctionals}. Two control strategies are derived in Section \ref{sec-combinedfunction}. Numerical results are given in Section \ref{sec-examples} followed by conclusions in Section \ref{sec-conclusion}.

\section{Preliminaries and Problem Formulation}
\label{sec-nonconsys}

Let $\mathbb{R}:=(-\infty, +\infty), \mathbb{R}_{+}:=[0, +\infty), \mathbb{N}:=\{0, 1, \ldots\}$ and $\mathbb{N}_{+}:=\{1, 2, \ldots\}$. For $x, y\in\mathbb{R}^{n}$, $(x, y):=[x^{\top}, y^{\top}]^{\top}$. $|\cdot|$ denotes the Euclidean norm. An open ball centered at $y\in\mathbb{R}^{n}$ with radius $\delta>0$ is denoted by $\mathbf{B}(y, \delta):=\{x\in\mathbb{R}^{n}: |x-y|<\delta\}$. $\mathbf{B}(\delta):=\mathbf{B}(0, \delta)$. $\mathcal{PC}([a, b], \mathbb{R}^{n})$ denotes the class of piecewise continuous functions from $[a, b]\subseteq\mathbb{R}$ to $\mathbb{R}^{n}$. $\mathcal{C}(\mathbb{R}^{n}, \mathbb{R}^{p})$ denotes the class of continuously differentiable functions mapping $\mathbb{R}^{n}$ to $\mathbb{R}^{p}$. For $x\in\mathcal{PC}([-\Delta, +\infty), \mathbb{R}^{n})$, let $x_{t}$ be an element of $\mathcal{PC}([-\Delta, 0], \mathbb{R}^{n})$ defined as $x_{t}(\theta):=x(t+\theta)$ with $t\in\mathbb{R}_{+}$ and $\theta\in[-\Delta, 0]$. $\|\phi\|:=\sup_{\theta\in[-\Delta, 0]}|\phi(\theta)|$ for any $\phi\in\mathcal{PC}([-\Delta, 0], \mathbb{R}^{n})$. The upper Dini derivative of a function $V\in\mathcal{C}(\mathbb{R}_{+}, \mathbb{R})$ is $D^{+}V(t):=\limsup_{s\rightarrow0^{+}}\frac{V(t+s)-V(t)}{s}$. For any $h: \mathcal{C}([-\Delta, 0], \mathbb{R}^{n})\rightarrow\mathbb{R}_{+}$, its upper Dini derivative is $D^{+}h(x_{t})=\limsup_{s\rightarrow0^{+}}\frac{h(x_{t+s})-h(x_{t})}{s}$. A continuous function $\alpha: \mathbb{R}_{+}\rightarrow\mathbb{R}_{+}$ is of class $\mathcal{K}$ if it is strictly increasing and $\alpha(0)=0$; it is of class $\mathcal{K}_{\infty}$ if it is of class $\mathcal{K}$ and unbounded. A continuous function $\beta: \mathbb{R}_{+}\times\mathbb{R}_{+}\rightarrow\mathbb{R}_{+}$ is of class $\mathcal{KL}$ if for each fixed $t\geq0$, $\beta(s, t)$ is of class $\mathcal{K}$, and for any fixed $s\geq0$, $\beta(s, t)$ decreases to 0 as $t\rightarrow\infty$.

\subsubsection{Interconnected Time-Delay Systems}
We consider the nonlinear interconnected time-delay system, which is denoted by $\mathcal{S}$ and has $p\in\mathbb{N}_{+}$ subsystems of the following dynamics:
\begin{align}
\label{eqn-1}
\mathcal{S}_{i}:\left\{\begin{aligned}
&\dot{x}_{i}=f_{i}(x_{t})+g_{i}(x_{t})u_{i}, &\quad&  t\geq0, \\
&x_{i}(t)=\xi_{i}(t), &\quad& t\in[-\Delta, 0],
\end{aligned}\right.
\end{align}
where $i\in\mathcal{N}:=\{1, \ldots, p\}$. For the $i$-th subsystem $\mathcal{S}_{i}$, $x_{i}\in\mathbb{R}^{n_{i}}$ is the state and $u_{i}\in\mathbb{R}^{m_{i}}$ is the control input. We denote by $x:=(x_{1}, \ldots, x_{p})\in\mathbb{R}^{n}$ the augmented state with $n:=\sum^{p}_{i=1}n_{i}$, and by $x_{t}:=(x^{1}_{t}, \ldots, x^{p}_{t})\in\mathcal{PC}([-\Delta, 0], \mathbb{R}^{n})$ the augmented time-delay state, where $\Delta>0$ is the upper bound of time delays. The initial condition is $\xi_{i}\in\mathcal{PC}([-\Delta, 0], \mathbb{X}_{i0})$ with $\mathbb{X}_{i0}\subset\mathbb{R}^{n_{i}}$ containing the origin. For all $i\in\mathcal{N}$, $\|\xi_{i}\|$ is assumed to be bounded. The functionals $f_{i}: \mathcal{PC}([-\Delta, 0], \mathbb{R}^{n})\rightarrow\mathbb{R}^{n_{i}}$ and $g_{i}: \mathcal{PC}([-\Delta, 0], \mathbb{R}^{n})\rightarrow\mathbb{R}^{n_{i}\times m_{i}}$ are assumed to be continuous and locally Lipschitz, which ensures the existence of the unique solution to the system \eqref{eqn-1}. Let $f_{i}(0)=0$ and $g_{i}(0)=0$. Hence, $x_{i}(t)\equiv0$ with $t>0$ is a trivial solution to each subsystem $\mathcal{S}_{i}$. 

For the system $\mathcal{S}$, the interconnection among all subsystems comes from the coupling and mutual communication, which can be characterized by a graph $\mathcal{G}:=\{\mathcal{N}, \mathcal{E}\}$ with the vertex set $\mathcal{N}$ and the edge set $\mathcal{E}\subseteq\mathcal{N}\times\mathcal{N}$. The graph $\mathcal{G}$ is assumed to be time-invariant and undirected. The time delays are from the spatial locations of all subsystems and remote communication. Both the interconnection relation and the time delays are embedded implicitly into the functions $f_{i}$ and $g_{i}$. That is, both $f_{i}$ and $g_{i}$ are rewritten as the functions of the time-delay state $x_{t}\in\mathcal{PC}([-\Delta, 0], \mathbb{R}^{n})$ but only depend on the time-delay states of the current subsystem and its neighbors. 

\begin{remark}
\label{rmk-1}
The dynamics \eqref{eqn-1} is called the \emph{control-affine} form and is general enough to model many physical systems like power systems \cite{Pham2018distributed}, transportation systems \cite{Giammarino2020traffic} and robotic systems \cite{Rodriguez2009bilateral}. We stress here that general nonlinear control systems can be transformed into nonlinear control-affine systems via many techniques like linearization techniques [32, Ch. 12] and backstepping techniques [37, Tab. 1]. Thus, it is of great importance to address the system \eqref{eqn-1}.  
\hfill $\square$
\end{remark}

\subsubsection{Stabilization Control}
The \textit{stabilization control} of interconnected time-delay systems is to design a distributed stabilizing feedback controller such that the closed-loop system is globally asymptotically stable, which is defined as follows.

\begin{definition}
\label{def-1}
Given all $u_{i}\in\mathbb{R}^{m_{i}}$ $i\in\mathcal{N}$, the system $\mathcal{S}$ is \emph{asymptotically stable (AS)}, if there exists $\beta\in\mathcal{KL}$ such that $|x(t)|\leq\beta(\|\xi\|, t)$ for all $t\geq0$ and $\xi\in\mathcal{PC}([-\Delta, 0], \mathbb{X}_{0})$ with $\mathbb{X}_{0}\subset\mathbb{R}^{n}$. The system $\mathcal{S}$ is \emph{globally asymptotically stable (GAS)}, if there exists $\beta\in\mathcal{KL}$ such that $|x(t)|\leq\beta(\|\xi\|, t)$ for all $t\geq0$ and all bounded $\xi\in\mathcal{PC}([-\Delta, 0], \mathbb{R}^{n})$.
\end{definition}

\begin{definition}[\cite{Di2017robustification}]
\label{def-2}
A functional $V: \mathcal{PC}([-\Delta, 0], \mathbb{R}^{n})\rightarrow\mathbb{R}_{+}$ is \emph{smoothly separable}, if there exist $V_{1}\in\mathcal{C}(\mathbb{R}^{n}, \mathbb{R}_{+})$, a locally Lipschitz functional $V_{2}: \mathcal{PC}([-\Delta, 0], \mathbb{R}^{n})\rightarrow\mathbb{R}_{+}$, and $\alpha_{1}, \alpha_{2}\in\mathcal{K}_{\infty}$ such that, for all $\phi\in\mathcal{PC}([-\Delta, 0], \mathbb{R}^{n})$,
\begin{align*}
V(\phi)&=V_{1}(\phi(0))+V_{2}(\phi), \\
\alpha_{1}(|\phi(0)|)&\leq V_{1}(\phi(0))\leq\alpha_{2}(|\phi(0)|).
\end{align*}
\end{definition}

\begin{definition}[\cite{Di2017robustification}]
\label{def-3}
A smoothly separable functional $V: \mathcal{PC}([-\Delta, 0], \mathbb{R}^{n})\rightarrow\mathbb{R}_{+}$ is \emph{invariantly differentiable (i-differentiable)}, if $V(\phi)=V_{1}(\phi(0))+V_{2}(\phi)$ and
\begin{enumerate}[(1)]
  \item for all $\phi\in\mathcal{PC}([-\Delta, 0], \mathbb{R}^{n})$ with $\mathbf{x}=\phi(0)$, both $\partial V_{1}(\mathbf{x})/\partial \mathbf{x}$ and $D^{+}V_{2}(\phi)$ exist;
  \item $D^{+}V_{2}(\phi)$ is invariant with respect to $\phi\in\mathcal{PC}([-\Delta, 0], \mathbb{R}^{n})$, that is, $D^{+}V_{2}(x_{0})$ is the same for all $x_{t}\in\mathcal{PC}([-\Delta, 0], \mathbb{R}^{n})$;
  \item  for all $x_{t}\in\mathcal{PC}([-\Delta, 0], \mathbb{R}^{n})$ and $l\geq0$, $V(x_{t+l})-V(x_{t}):=\frac{\partial V_{1}(y)}{\partial y}z+D^{+}V_{2}(x_{t})l+o(\sqrt{|z|^{2}+l^{2}})$, where $y=x_{t}(0), z=x_{t+l}(0)-x_{t}(0)$ and $\lim_{s\rightarrow0^{+}}o(s)/s=0$.
\end{enumerate}
In addition, $V$ is \emph{continuously i-differentiable} if $D^{+}V_{2}(\phi)$ is continuous.
\end{definition}

\subsubsection{Safety Control}
The \textit{safety control} of interconnected time-delay systems is to design a distributed controller such that all system states stay in a predefined set. For each $\mathcal{S}_{i}$ with $i\in\mathcal{N}$, a set $\mathbb{A}_{i}\subset\mathbb{R}^{n_{i}}$ is \emph{forward invariant}, if $x_{i}(t)\in\mathbb{A}_{i}$ for any trajectory $x_{i}(t)$ starting from $\xi_{i}\in\mathcal{PC}([-\Delta, 0], \mathbb{A}_{i})$. If the set $\mathbb{A}_{i}$ is forward invariant, then the subsystem $\mathcal{S}_{i}$ is \emph{safe} with respect to $\mathbb{A}_{i}$; and the set $\mathbb{A}_{i}$ is called the \emph{safe set}. 

To address the safety control problem, each $\mathcal{S}_{i}$ has a safe set $\mathbb{S}_{i}\subset\mathcal{PC}([-\Delta, 0], \mathbb{R}^{n_{i}})$, which is associated with a continuously differential functional $h_{i}: \mathcal{PC}([-\Delta, 0], \mathbb{R}^{n})\rightarrow\mathbb{R}$. 
\begin{align}
\label{eqn-2}
\mathbb{S}_{i}&:=\{\phi_{i}\in\mathcal{PC}([-\Delta, 0], \mathbb{R}^{n_{i}}): h_{i}(\phi)\geq0\}, \\
\label{eqn-3}
\partial\mathbb{S}_{i}&:=\{\phi_{i}\in\mathcal{PC}([-\Delta, 0], \mathbb{R}^{n_{i}}): h_{i}(\phi)=0\}, \\
\label{eqn-4}
\inte(\mathbb{S}_{i})&:=\{\phi_{i}\in\mathcal{PC}([-\Delta, 0], \mathbb{R}^{n_{i}}): h_{i}(\phi)>0\}.
\end{align}
Let $\inte(\mathbb{S}_{i})\neq\varnothing$ and $\overbar{\inte(\mathbb{S}_{i})}=\mathbb{S}_{i}$. For each subsystem, $\mathbb{S}_{i}$ is forward invariant if $x^{i}_{t}\in\mathbb{S}_{i}$ for all $t\geq0$. Let $\mathbb{S}:=\prod^{i}_{i=1}\mathbb{S}_{i}$. 

Note that the functional $h_{i}$ in \eqref{eqn-2}-\eqref{eqn-4} is on $\mathcal{PC}([-\Delta, 0], \mathbb{R}^{n})$ such that all time-delay states are involved, which is different from many existing works \cite{Lyu2022small}. This setting is reasonable since the safe set of each subsystem is inevitably related to the neighbor subsystems such that the collisions among all subsystems can be avoided; see, e.g., \cite{Wang2017safety, Panagou2015distributed} for the delay-free cases. If $h_{i}$ is defined on $\mathcal{PC}([-\Delta, 0], \mathbb{R}^{n_{i}})$, then the safe set of each subsystem does not depend on the neighbor subsystems, which is a special case of our setting.

\section{Multiple Control Functionals}
\label{sec-multiplefunctionals}

In this section multiple control Lyapunov and barrier functionals are proposed respectively for the stabilization and safety control problems of interconnected time-delay systems.

\subsection{Multiple Control Lyapunov Functionals}

\begin{definition}
\label{def-4}
For the system $\mathcal{S}$, the continuously i-differentiable functionals $V_{i}: \mathcal{PC}([-\Delta, 0], \mathbb{R}^{n_{i}})\rightarrow\mathbb{R}_{+}$ are called the \textit{multiple control Lyapunov functionals (MCLFs)}, if 
\begin{enumerate}[(i)]
  \item for all $\phi_{i}\in\mathcal{PC}([-\Delta, 0], \mathbb{R}^{n_{i}}), i\in\mathcal{N}$, $\alpha_{i1}, \alpha_{i2}\in\mathcal{K}_{\infty}$ exist such that $\alpha_{i1}(|\phi_{i}(0)|)\leq V_{i}(\|\phi_{i}\|)\leq\alpha_{i2}(\|\phi_{i}\|)$; 
  
  \item for all $i, j\in\mathcal{N}$, there exist $\rho_{i}, \gamma_{ij}\in\mathcal{K}$ such that for all $\phi_{i}\in\mathcal{PC}([-\Delta, 0], \mathbb{R}^{n_{i}})$,
  \begin{align}
  \label{eqn-5}
  &\inf\nolimits_{u_{i}\in\mathbb{R}^{m_{i}}}\{L_{f_{i}}V_{i1}(\phi)+D^{+}V_{i2}(\phi_{i})+L_{g_{i}}V_{i1}(\phi)u_{i}\} \nonumber \\
  &\quad <-\rho_{i}(V_{i}(\phi_{i}))+\sum\nolimits_{j\in\mathcal{N}}\gamma_{ij}(V_{j}(\phi_{j})), 
  \end{align}
  where $L_{f_{i}}V_{i1}(\phi):=\frac{\partial V_{i1}(\phi_{i}(0))}{\partial \phi_{i}(0)}f_{i}(\phi)$ and $L_{g_{i}}V_{i1}(\phi):=\frac{\partial V_{i1}(\phi_{i}(0))}{\partial \phi_{i}(0)}g_{i}(\phi)$;
  
  \item for all $i\in\mathcal{N}$ and all nonzero $s\in\mathcal{PC}([-\Delta, 0], \mathbb{R}_{+}^{n})$, there exist bounded positive definite functions $\zeta_{i}: \mathbb{R}\rightarrow\mathbb{R}_{+}$ such that $\int^{\infty}_{0}\zeta_{i}(\rho_{i}(s))ds=\infty$ and $\zeta^{\top}(s)\Gamma_{1}(A^{-1}(s))<\zeta^{\top}(s)s$, where $\zeta=(\zeta_1, \ldots, \zeta_{p})$, $\Gamma_{1}(s)=(\sum_{j\neq1}\gamma_{1j}(s_{j}),\ldots, \sum_{j\neq p}\gamma_{pj}(s_{j}))$ and $A(s)=(\rho_{1}(s_{1}), \ldots, \rho_{p}(s_{p}))$. 
\end{enumerate}
\end{definition}

\begin{remark}
\label{rmk-2}
From Definition \ref{def-4}, each subsystem admits a Lyapunov-like function such that an ISS-like condition (i.e., \eqref{eqn-5}) is satisfied in a dissipative manner. That is, for each subsystem, its neighbor subsystems are treated as the external disturbances; see the second term of the right-hand side of \eqref{eqn-5}. Item (iii) is the constraint on the functions $\rho_{i}, \gamma_{ij}$ and plays a similar role as the small-gain condition; see also \cite{Dashkovskiy2011small}. In addition, the proposed MCLFs are the so-called Krasovskii version, which can be reduced to the Razumikhin version by setting $V_{i}(\phi_{i})=V_{i1}(\phi_{i}(0))$, which relates to the current state only. For this case, the follow-up analysis is still valid via a slight modification and hence is omitted here. 
\hfill $\square$
\end{remark}

\begin{definition}
\label{def-5}
The system $\mathcal{S}$ is said to satisfy the \textit{distributed small control property (DSCP)}, if for each $\varepsilon_{i}>0$, $i\in\mathcal{N}$, there exist $\delta_{i}>0$ such that for all nonzero $\phi_{i}\in\mathcal{PC}([-\Delta, 0], \mathbf{B}(\delta_{i}))$, there exists $u_{i}\in\mathbf{B}(\varepsilon_{i})$ such that $L_{f_{i}}V_{i1}(\phi)+D^{+}V_{i2}(\phi_{i})+L_{g_{i}}V_{i1}(\phi)u_{i}<-\rho_{i}(V_{i}(\phi_{i}))+\sum_{j\in\mathcal{N}}\gamma_{ij}(V_{j}(\phi_{j}))$. 
\end{definition}

Definitions \ref{def-4} and \ref{def-5} extend the classic versions in \cite{Sontag1989universal} and the time-delay versions in \cite{Ren2022Razumikhin2} to the interconnected time-delay case. As a result, the distributed controller can be derived explicitly such that the closed-loop system is GAS.

\begin{theorem}
\label{thm-1}
If the system $\mathcal{S}$ admits MCLFs and satisfies the DSCP, then the closed-loop system is GAS under the continuous controller designed below:
\begin{align}
\label{eqn-6}
u_{i}(\phi)=\left\{\begin{aligned}
&\kappa_{i}(\mathfrak{a}_{i}(\phi), \mathfrak{b}^{\top}_{i}(\phi)), &&\text{if }  \phi\neq0\wedge\mathfrak{b}_{i}(\phi)\neq0, \\
&0, &&\text{otherwise}, 
\end{aligned}\right.
\end{align}
where $i\in\mathcal{N}$, $\mathfrak{a}_{i}(\phi):=L_{f_{i}}V_{i1}(\phi)+D^{+}V_{i2}(\phi_{i})+\rho_{i}(V_{i}(\phi_{i}))-\sum_{j\in\mathcal{N}}\gamma_{ij}(V_{j}(\phi_{j}))$, $\mathfrak{b}_{i}(\phi):=L_{g_{i}}V_{i1}(\phi)$, and
\begin{equation*}
\kappa_{i}(\mathfrak{a}_{i}(\phi), \mathfrak{b}_{i}(\phi))=\frac{\mathfrak{a}_{i}(\phi)+\sqrt{\mathfrak{a}^{2}_{i}(\phi)+\|\mathfrak{b}_{i}(\phi)\|^{4}}}{-\|\mathfrak{b}_{i}(\phi)\|^{2}}\mathfrak{b}_{i}(\phi).
\end{equation*}
\end{theorem}

\begin{IEEEproof}
From \eqref{eqn-6}, if either $\phi=0$ or $\mathfrak{b}_{i}(\phi)=L_{g_{i}}V_{i}(\phi)=0$, then $u_{i}\equiv0$. In this case, from Definition \ref{def-4}, we have $L_{f_{i}}V_{i1}(\phi)+D^{+}V_{i2}(\phi_{i})<-\rho_{i}(V_{i}(\phi_{i}))+\sum_{j\in\mathcal{N}}\gamma_{ij}(V_{j}(\phi_{j}))$. If $\phi\neq0$ and $\mathfrak{b}_{i}(\phi)\neq0$, then
\begin{align*}
&L_{f_{i}}V_{i1}(\phi)+D^{+}V_{i2}(\phi_{i})+L_{g_{i}}V_{i}(\phi)u_{i} \\
&=-\sqrt{\mathfrak{a}^{2}_{i}(\phi)+\|\mathfrak{b}_{i}(\phi)\|^{4}}-\rho_{i}(V_{i}(\phi_{i}))+\sum\nolimits_{j\in\mathcal{N}}\gamma_{ij}(V_{j}(\phi_{j})) \\
&\leq-\rho_{i}(V_{i}(\phi_{i}))+\sum\nolimits_{j\in\mathcal{N}}\gamma_{ij}(V_{j}(\phi_{j})). 
\end{align*}
Hence, under the controller \eqref{eqn-6}, we have 
\begin{align}
\label{eqn-7}
D^{+}V_{i}(\phi_{i})&=L_{f_{i}}V_{i1}(\phi)+D^{+}V_{i2}(\phi_{i})+L_{g_{i}}V_{i}(\phi)u_{i} \nonumber \\
&\leq-\rho_{i}(V_{i}(\phi_{i}))+\sum\nolimits_{j\in\mathcal{N}}\gamma_{ij}(V_{j}(\phi_{j})). 
\end{align}
Since item (iii) holds, we conclude from \eqref{eqn-7} and \cite[Thm. 4.1]{Dashkovskiy2011small} that the system $\mathcal{S}$ is GAS.

If $\phi\neq0$, then the continuity of the controller \eqref{eqn-6} comes from the continuity of $\mathfrak{a}_{i}(\phi)$ and $\mathfrak{b}_{i}(\phi)$. Next we only consider the continuity of \eqref{eqn-6} at the origin. First, from the DSCP in Definition \ref{def-5}, for arbitrary $\varepsilon_{i}>0$, there exists $\delta_{i}>0$ such that for any nonzero $\phi_{i}\in\mathcal{PC}([-\Delta, 0], \mathbf{B}(\delta_{i}))$, there exists $u_{i}\in\mathbf{B}(\varepsilon_{i})$ such that $\mathfrak{a}_{i}(\phi)+\mathfrak{b}_{i}(\phi)u_{i}<0$. Second, since $V_{i1}\in\mathcal{C}(\mathbb{R}^{n}, \mathbb{R}_{+})$ and $g_{i}$ in \eqref{eqn-1} is locally Lipschitz, there exists $\bar{\delta}_{i}>0$ with $\bar{\delta}_{i}\neq\delta_{i}$ such that $\|\mathfrak{b}_{i}(\phi)\|\leq\varepsilon_{i}$ holds for all nonzero $\phi_{i}\in\mathcal{PC}([-\Delta, 0], \mathbf{B}(\bar{\delta}_{i}))$. Let $\tilde{\delta}_{i}:=\min\{\delta_{i}, \bar{\delta}_{i}\}$. Finally, for any nonzero $\phi_{i}\in\mathcal{PC}([-\Delta, 0], \mathbf{B}(\tilde{\delta}_{i}))$, $\|\mathfrak{b}_{i}(\phi)\|\leq\varepsilon_{i}$ and there exists $u_{i}\in\mathbf{B}(\varepsilon_{i})$ such that $\mathfrak{a}_{i}(\phi)+\mathfrak{b}_{i}(\phi)u_{i}<0$. With the above analysis, two cases are discussed below. 

If $\mathfrak{b}_{i}(\phi)=0$, then $u_{i}(\phi)=0$ from \eqref{eqn-6}. Since $u_{i}(0)=0$ and $\varepsilon_{i}\in\mathbb{R}_{+}$ can be arbitrarily small, the controller \eqref{eqn-6} is continuous at the origin. If $\mathfrak{b}_{i}(\phi)\neq0$, then  $\|\mathfrak{a}_{i}(\phi)\|\leq\varepsilon_{i}\|\mathfrak{b}_{i}(\phi)\|$ for any nonzero $\phi_{i}\in\mathcal{PC}([-\Delta, 0], \mathbf{B}(\tilde{\delta}_{i}))$. In this case, for any nonzero $\phi_{i}\in\mathcal{PC}([-\Delta, 0], \mathbf{B}(\tilde{\delta}_{i}))$,
\begin{align*}
\|u_{i}(\phi)\|&\leq\left|\frac{\mathfrak{a}_{i}(\phi)+\sqrt{\mathfrak{a}^{2}_{i}(\phi)+\|\mathfrak{b}_{i}(\phi)\|^{4}}}{\|\mathfrak{b}_{i}(\phi)\|}\right|  \\
&\leq\left|\frac{\mathfrak{a}_{i}(\phi)+\|\mathfrak{a}_{i}(\phi)\|+\|\mathfrak{b}_{i}(\phi)\|^{2}}{\|\mathfrak{b}_{i}(\phi)\|}\right|\leq3\varepsilon_{i}.
\end{align*}
Since $\varepsilon_{i}\in\mathbb{R}_{+}$ can be arbitrarily small, the controller \eqref{eqn-6} is continuous at the origin. Summarizing all above analysis, we conclude that the controller \eqref{eqn-6} is indeed continuous.
\end{IEEEproof}

Theorem \ref{thm-1} shows how to design the distributed controller to ensure the GAS property of interconnected time-delay systems. It is easy to check that the controller of each subsystem involves the states from its neighbor subsystems, and thus the controllers of all subsystems are coupling with each other. 

\subsection{Multiple Control Barrier Functionals}
\label{sec-vectorbarrier}

To investigate the safety of the system $\mathcal{S}$, multiple control barrier functionals are proposed in this subsection. 

\begin{definition}
\label{def-6}
For the system $\mathcal{S}$, the continuously i-differentiable functionals $B_{i}: \mathcal{PC}([-\Delta, 0], \mathbb{R}^{n})\rightarrow\mathbb{R}$ are called the \textit{multiple control barrier functionals (MCBFs)}, if
\begin{enumerate}[(i)]
  \item for all $i\in\mathcal{N}$, there exist $\alpha_{i1}, \alpha_{i2}\in\mathcal{K}_{\infty}$ such that for all $\phi\in\mathcal{PC}([-\Delta, 0], \mathbb{R}^{n})$, $\alpha_{i1}(h_{i}(\phi))\leq1/B_{i}(\phi)\leq\alpha_{i2}(h_{i}(\phi))$, where $h_{i}$ is defined in \eqref{eqn-2};
  
  \item for all $i, j\in\mathcal{N}$ there exist $\eta_{i}, \chi_{ij}\in\mathcal{K}$ such that for all $\phi\in\mathcal{PC}([-\Delta, 0], \mathbb{R}^{n})$,
  \begin{align}
  \label{eqn-8}
  &\inf\nolimits_{u_{i}\in\mathbb{R}^{m_{i}}}\{L_{f}B_{i1}(\phi)+D^{+}B_{i2}(\phi)+L_{g_{i}}B_{i1}(\phi)u_{i}\} \nonumber \\
  &\quad <\eta_{i}(h_{i}(\phi))-\sum\nolimits_{j\in\mathcal{N}}\chi_{ij}(h_{j}(\phi)), 
  \end{align}
  where $L_{f}B_{i1}(\phi):=\sum_{j\in\mathcal{N}}\frac{\partial B_{i1}(\phi(0))}{\partial \phi_{j}(0)}f_{j}(\phi)$ and $L_{g_{i}}B_{i1}(\phi):=\frac{\partial B_{i1}(\phi(0))}{\partial\phi_{_{i}}(0)}g_{i}(\phi)$;
  
  \item for all $i\in\mathcal{N}$ and all nonzero $s\in\mathcal{PC}([-\Delta, 0], \mathbb{R}_{+}^{n})$, there exist bounded positive definite functions $\omega_{i}: \mathbb{R}\rightarrow\mathbb{R}_{+}$ such that $\int^{\infty}_{0}\omega_{i}(\rho_{i}(s))ds=\infty$ and $\omega^{\top}(s)\Gamma_{2}(H^{-1}(s))<\omega^{\top}(s)s$, where $\omega=(\omega_{1}, \ldots, \omega_{p})$, $\Gamma_{2}(s)=(\sum_{j\neq1}\chi_{1j}(s), \ldots, \sum_{j\neq p}\chi_{pj}(s))$ and $H(s)=(\eta_{1}(s), \ldots, \eta_{p}(s))$. 
\end{enumerate}
\end{definition}

Definition \ref{def-6} extends the existing control barrier functions \cite{Ren2022Razumikhin, Ren2022Razumikhin2} to the interconnected time-delay case. In \eqref{eqn-8}, only the term $L_{g_{i}}B_{i1}(\phi)u_{i}$ is involved, which is a distributed version to avoid the reliance on a centralized coordination strategy. Note that the centralized version is $L_{g}B_{i1}(\phi)u$ with $L_{g}B_{i1}(\phi):=(\frac{\partial B_{i1}(\phi(0))}{\partial \phi_{1}(0)}g_{1}(\phi), \ldots, \frac{\partial B_{i1}(\phi(0))}{\partial \phi_{p}(0)}g_{p}(\phi))$. If the number $p$ is large, then the centralized version may encounter huge computation burden so that it is hard to be implemented. The distributed version \eqref{eqn-8} is for each subsystem, and later we will show that the safety guarantee is still valid. For the delay-free case, the reasonability of such a distributed version has been discussed in \cite{Wang2016safety} for multi-robot systems. 

\begin{theorem}
\label{thm-2}
Consider the system $\mathcal{S}$ admitting the MCBFs $B_{i}: \inte(\mathbb{S}_{i})\rightarrow\mathbb{R}$ with $\mathbb{S}_{i}$ in \eqref{eqn-2}. The set $\inte(\mathbb{S})$ is forward invariant under a Lipschitz continuous controller $u_{i}\in\mathbb{K}_{i}$ with
$\mathbb{K}_{i}:=\{u_{i}\in\mathbb{U}_{i}: L_{f}B_{i1}(\phi)+D^{+}B_{i2}(\phi)+L_{g_{i}}B_{i1}(\phi)u_{i}<\eta_{i}(h_{i}(\phi))-\sum\nolimits_{j\in\mathcal{N}}\chi_{ij}(h_{j}(\phi))\}$.
\end{theorem}

\begin{IEEEproof}
For the MCBFs $B_{i}: \mathcal{PC}([-\Delta, 0], \mathbb{R}^{n})\rightarrow\mathbb{R}$, $i\in\mathcal{N}$, we define the functional $\mathcal{B}_{i}(t):=1/B_{i}(x_t)$ and further have $D^{+}\mathcal{B}_{i}(\phi)=-D^{+}B_{i}(\phi)/(B^{2}_{i}(\phi))$ and 
\begin{align}
\label{eqn-9}
D^{+}\mathcal{B}_{i}(\phi)&=-\mathcal{B}^{2}_{i}(\phi)\left(L_{f}B_{i1}(\phi)+D^{+}B_{i2}(\phi)+L_{g_{i}}B_{i1}(\phi)u_{i}\right)\nonumber\\
&>-\mathcal{B}^{2}_{i}(\phi)\left(\eta_{i}(h_{i}(\phi))-\sum\nolimits_{j\in\mathcal{N}}\chi_{ij}(h_{j}(\phi))\right) \nonumber \\
&\geq-\bar{\eta}_{i}(\mathcal{B}_{i}(\phi))+\sum\nolimits_{j\in\mathcal{N}}\bar{\chi}_{ij}(\mathcal{B}_{j}(\phi)), 
\end{align}
where $\bar{\eta}_{i}(\mathcal{B}_{i}(\phi)):=\mathcal{B}_{i}^{2}(\phi)\eta_{i}(\alpha^{-1}_{i2}(\mathcal{B}_{i}(\phi)))$ and $\bar{\chi}_{ij}(\mathcal{B}_{j}(\phi)):=\mathcal{B}_{i}^{2}(\phi)\chi_{ij}(\alpha^{-1}_{1}(\mathcal{B}_{i}(\phi)))$, both of which are of class $\mathcal{K}$. From \eqref{eqn-9}, \cite[Thm. 3.1]{Pepe2006lyapunov} and the comparison principle, there exists $\zeta_{i}\in\mathcal{KL}$ such that
\begin{align*}
\mathcal{B}_{i}(t)\geq\zeta_{i}(\mathcal{B}_{i}(0), t), \quad  \forall t\geq0, 
\end{align*}
combining which with the definition of $\mathcal{B}_{i}(\phi)$ yields that for all $t\geq0$, $1/B_{i}(x_{t})\geq\zeta_{i}(1/B_{i}(x_{0}), t)$, which further implies from item (i) of Definition \ref{def-6} that
\begin{equation}
\label{eqn-10}
\alpha_{i2}(h_{i}(x_{t}))\geq\zeta_{i}(\alpha_{i1}(h_{i}(\|\xi\|)), t), \quad  \forall t\geq0.
\end{equation}
Hence, from \eqref{eqn-10}, it is easy to check that $\alpha_{i2}(h_{i}(x_{t}))>0$ for all $t\geq0$. That is,  $h_{i}(x_{t})$ for all $t\geq0$, and the set $\inte(\mathbb{S})$ is forward invariant, which completes the proof.
\end{IEEEproof}

\begin{remark}
\label{rmk-3}
The proposed multiple control functionals are for interconnected time-delay systems and extend many existing works like \cite{Ren2022Razumikhin2, Ren2023vector, Lyu2022small}. In addition, different from \cite{Dashkovskiy2011small, Pepe2006lyapunov} on stability analysis only, a further step is made here to address the stabilization and safety control problems. 
\hfill $\square$
\end{remark}

\section{Safe Stabilization Controller Design}
\label{sec-combinedfunction}

With the proposed multiple control functionals, in this section we address the safe stabilization control problem, which aims to design a distributed controller such that the stabilization and safety can be guaranteed simultaneously. 

\subsubsection{Optimization-based Design}
For the system $\mathcal{S}$, let the MCLFs be $V_{i}$ and the MCBFs be $B_{i}$. Since the distributed stabilizing controller can be designed explicitly in Theorem \ref{thm-1}, the next is how to guarantee the safety objective proposed in Definition \ref{def-6}. To resolve this problem, we formulate the following optimization problem for each subsystem:
\begin{align}
\label{eqn-11}
\begin{aligned}
\min&\quad  \|u_{i}-u^{i}_{\mathsf{nom}}\|^{2} \\
\text{s.t.}&\quad A_{i}u_{i}\leq b_{i}, \quad i\in\mathcal{N}, 
\end{aligned}
\end{align}
where $u^{i}_{\mathsf{nom}}$ is the controller \eqref{eqn-6}, $A_{i}:=L_{g_{i}}B_{i1}(\phi)$ and $b_{i}:=L_{f}B_{i1}(\phi)+D^{+}B_{i2}(\phi)-\eta_{i}(B_{i}(\phi))+\sum_{j\in\mathcal{N}}\chi_{ij}(B_{j}(\phi))$. The problem \eqref{eqn-11} is a distributed quadratic programming (QP) problem, whose solution is a QP-based controller. In \eqref{eqn-11}, $u^{i}_{\mathsf{nom}}$ is treated as the nominal controller, and the resulting controller is to modify the nominal controller when the safety becomes imminent. Note that the constraint in \eqref{eqn-11} is of the half-plane form, since the MCBFs exist and $A_{i}, b_{i}$ can be computed explicitly. Hence, \eqref{eqn-11} can be solved via many existing tools. In addition, if the control inputs are constrained, then the input constraints can be embedded into \eqref{eqn-11}. 

\subsubsection{Sliding Mode based Design}
An alternative way to combine the MCLFs and MCBFs is via multiple sliding surface functionals, which are defined as 
\begin{align}
\label{eqn-12}
U_{i}(\phi)&:=\psi_{i}(V_{i}(\phi_{i}), B_{i}(\phi)).
\end{align}
Let $U_{i}: \mathcal{PC}([-\Delta, 0], \mathbb{R}^{n})\rightarrow\mathbb{R}$ and $\psi_{i}: \mathbb{R}\times\mathbb{R}\rightarrow\mathbb{R}$ be continuously differentiable and radially unbounded. Hence,
\begin{equation}
\label{eqn-13}
D^{+}U_{i}(\phi):=\mathbf{F}_{i}(\phi)+\mathbf{G}_{i}(\phi)u_{i}+\mathbf{L}_{i}(\phi)
\end{equation}
with $\mathbf{F}_{i}(\phi):=\mathbf{H}_{i}(\phi)f_{i}(\phi), \mathbf{G}_{i}(\phi):=\mathbf{H}_{i}(\phi)g_{i}(\phi)$ and
\begin{align*}
\mathbf{H}_{i}(\phi)&:=\frac{\partial\psi_{i}}{\partial{V_{i}}}\frac{\partial{V}_{i1}(\phi_{i}(0))}{\partial\phi_{i}(0)}+\frac{\partial\psi_{i}}{\partial{B_{i}}}\frac{\partial{B}_{i1}(\phi_{i}(0))}{\partial\phi_{i}(0)}, \\ \mathbf{L}_{i}(\phi)&:=\frac{\partial\psi_{i}}{\partial{V_{i}}}D^{+}{V}_{i2}(\phi)+\frac{\partial\psi_{i}}{\partial{B_{i}}}D^{+}{B}_{i2}(\phi).
\end{align*}

Based on \eqref{eqn-13}, two auxiliary functionals are introduced:
\begin{align*}
\mathbf{J}_{i1}(\phi)&:=\frac{g_{i}(\phi)\mathbf{G}^{\top}_{i}(\phi)f^{\top}_{i}(\phi)-f_{i}(\phi)\mathbf{G}_{i}(\phi)g^{\top}_{i}(\phi)}{2\|\mathbf{G}_{i}(\phi)\|^{2}}, \\
\mathbf{J}_{i2}(\phi)&:=\frac{g_{i}(\phi)\mathbf{G}^{\top}_{i}(\phi)f^{\top}_{i}(\phi)+f_{i}(\phi)\mathbf{G}_{i}(\phi)g^{\top}_{i}(\phi)}{2\|\mathbf{G}_{i}(\phi)\|^{2}}.
\end{align*}
Let $\mathbf{G}_{i}(\phi)\neq0$. If $\mathbf{G}_{i}(\phi)=0$, then higher-order sliding surface functionals can be introduced to ensure that the follow-up analysis can be proceeded similarly \cite{Ren2022Razumikhin2}. Since $\mathbf{H}_{i}(\phi)f_{i}(\phi)\in\mathbb{R}$ and $\mathbf{H}_{i}(\phi)g_{i}(\phi)g^{\top}_{i}(\phi)\mathbf{H}^{\top}_{i}(\phi)$ is symmetric, we can check that $\mathbf{H}_{i}(\phi)\mathbf{J}_{i1}(\phi)\mathbf{H}^{\top}_{i}(\phi)=0$ and $\mathbf{H}_{i}(\phi)(\mathbf{J}_{i1}(\phi)+\mathbf{J}_{i2}(\phi))\mathbf{H}^{\top}_{i}(\phi)=\mathbf{F}_{i}(\phi)$. In the ideal case, the system trajectory is expected to satisfy the manifold invariant condition $U_{i}(\phi)=0$, which can be verified via the functional $W_{i}(\phi):=0.5U^{2}_{i}(\phi)$. More specifically, 
\begin{align}
\label{eqn-14}
D^{+}W_{i}(\phi)&=U_{i}(\phi)D^{+}U_{i}(\phi)\nonumber\\
& =U_{i}(\phi)(-\mathbf{H}_{i}(\phi)\mathbf{J}_{i1}(\phi)\mathbf{H}^{\top}_{i}(\phi)\nonumber\\
&\quad  +\mathbf{H}_{i}(\phi)\mathbf{J}_{i2}(\phi)\mathbf{H}^{\top}_{i}(\phi)+\mathbf{L}_{i}(\phi)+\mathbf{G}_{i}(\phi)u_{i}) \nonumber\\
& =U_{i}(\phi)(\mathbf{H}_{i}(\phi)\mathbf{J}_{i2}(\phi)\mathbf{H}^{\top}_{i}(\phi) \nonumber\\
&\quad  +\mathbf{L}_{i}(\phi)+\mathbf{G}_{i}(\phi)u_{i}).
\end{align}
Let $D^{+}W_{i}(\phi)=0$, and the ideal controller is designed as
\begin{align*}
\bar{u}_{i}(\phi):=-\|\mathbf{G}_{i}(\phi)\|^{-2}\mathbf{G}^{\top}_{i}(\phi)(\mathbf{H}_{i}(\phi)\mathbf{J}_{i2}(\phi)\mathbf{H}^{\top}_{i}(\phi)+\mathbf{L}_{i}(\phi)).
\end{align*}
Since the exact state of the system $\mathcal{S}$ may move into the sublevel and superlevel sets of the sliding surface, the applied controller is not the same as $\bar{u}_{i}$ but modified as 
\begin{align}
\label{eqn-15}
u_{i}(\phi):=\bar{u}_{i}(\phi)-\|\mathbf{G}_{i}(\phi)\|^{-2}\mathbf{G}^{\top}_{i}(\phi)\mathbf{K}_{i}(\phi),
\end{align}
where $i\in\mathcal{N}$ and $\mathbf{K}_{i}(\phi)$ is an additional item to be designed. From all above discussion, we derive the following theorem, which guarantees simultaneously the stabilization and safety objectives for the system $\mathcal{S}$ via $U_{i}(\phi)$ in \eqref{eqn-12}.

\begin{theorem}
\label{thm-3}
Consider the system $\mathcal{S}$ with the safe set $\mathbb{S}_{i}\subset\mathcal{PC}([-\Delta, 0], \mathbb{R}^{n_{i}})$ in \eqref{eqn-2}-\eqref{eqn-4}. Let $\xi_{i}\in\inte(\mathbb{S}_{i})$. If the functional $U_{i}$ in \eqref{eqn-12} is such that
\begin{align}
\label{eqn-16}
U^{2}_{i}(\phi)&\geq U^{2}_{i}(\xi), \quad \forall \phi\in\partial\mathbb{S}, \\
\label{eqn-17}
\mathbb{A}_{i}&:=\{\phi_{i}\in\mathbb{S}_{i}: U_{i}(\phi)=0\}\subset\inte(\mathbb{S}_{i}),
\end{align}
then the stabilization and safety objectives can be achieved simultaneously via the controller \eqref{eqn-15} with $\mathbf{K}_{i}(\phi):=\mathsf{K}_{i}U_{i}(\phi)/(\|U_{i}(\phi)\|+\varpi_{i})$, where $\mathsf{K}_{i}>0$ is constant and $\varpi_{i}>0$ is sufficiently small.
\end{theorem}

\begin{IEEEproof}
From \eqref{eqn-14} and \eqref{eqn-15}, we have
\begin{align*}
D^{+}W_{i}(\phi)&\leq U_{i}(\phi)(\mathbf{H}_{i}(\phi)\mathbf{J}_{i2}(\phi)\mathbf{H}^{\top}_{i}(\phi)+\mathbf{L}_{i}(\phi)+\mathbf{G}_{i}(\phi)u_{i})  \nonumber  \\
&=-\frac{\mathsf{K}_{i}U^{2}_{i}(\phi)}{\|U_{i}(\phi)\|+\varpi_{i}}=:-\mathsf{K}_{i}\eta_{i}(\phi),
\end{align*}
where $\eta_{i}(\phi):=\frac{U^{2}_{i}(\phi)}{\|U_{i}(\phi)\|+\varpi_{i}}$. From \cite[Thm. 3.1]{Pepe2006lyapunov}, $W_{i}(\phi)$ is monotonically decreasing and converges to the origin asymptotically, which implies the satisfaction of the stabilization objective. From \eqref{eqn-17} and the manifold invariant condition, the sliding surface is in the safe set. From \eqref{eqn-16}, we have $|U_{i}(\phi(\theta))|\geq|U_{i}(\xi(\theta))|$ for all $\theta\in[-\Delta, 0]$. Hence, from the convergence of the functional $W_{i}(t)$, the state trajectory starting from the initial condition is convergent along the sliding surface, while avoiding to cross the boundary of the safe set. Therefore, the safety objective is guaranteed.
\end{IEEEproof}

\begin{remark}
\label{rmk-4}
The proposed two control methods have their own advantages in dealing with specific control problems. The optimization-based method is to formulate a distributed QP problem to be solved in real time and can be further extended to deal with the input saturation problem. The sliding mode based method applies the sliding surface functionals to propose an explicit way to design the closed-form distributed controller. However, how to solve the time-delay optimization problems efficiently and how to construct the sliding surface functionals depend on the sizes of the considered systems and tasks, and deserve further study. 
\hfill $\square$
\end{remark}

\section{Numerical Results}
\label{sec-examples}

Consider four omnidirectional robots, whose states are $x_{i}:=(x_{i1}, x_{i2}, x_{i3})\in\mathbb{R}^{3}$ with $i\in\{1, 2, 3, 4\}$. To be specific, $\mathfrak{p}_{i}:=(x_{i1}, x_{i2})\in\mathbb{R}^{2}$ is the robot position, and $x_{i3}\in\mathbb{R}$ is the robot orientation with respect to $x_{i1}$. Let $x:=(x_{1}, x_{2}, x_{3}, x_{4})$ be the stacked states. The time-delay coupling among all robots is denoted as the function $f(x_{t}):=(f_{1}(x_{t}), f_{2}(x_{t}), f_{3}(x_{t}), f_{4}(x_{t}))\in\mathbb{R}^{12}$ with $f_{i}(x_{t})=(f_{i1}(x_{t}), f_{i2}(x_{t}), 0)$, and
\begin{align*}
f_{il}(x_{t})&:=\sum\nolimits^{4}_{j=1}\frac{k_{i}(x^{il}_{t}-x^{jl}_{t})}{\|\mathfrak{p}^{i}_{t}-\mathfrak{p}^{j}_{t}\|+\varepsilon_{i}}, \quad k_{i}>0, \quad l\in\{1, 2\},
\end{align*}
where $x^{i}_{t}=(x^{i1}_{t}, x^{i2}_{t}, x^{i3}_{t})$ is the time-delay state, $\mathfrak{p}^{i}_{t}=(x^{i1}_{t}, x^{i2}_{t})$ and $\varepsilon_{i}>0$ is arbitrarily small. The upper bound of the time delays is set as 0.5.  Hence, $f(x)$ is locally Lipschitz continuous, and the dynamics for each robot is 
\begin{align}
\label{eqn-18}
\dot{x}_{i} &=f_{i}(x_{t})+\begin{bmatrix}
\cos(x_{i3}) & -\sin(x_{i3}) & 0 \\  \sin(x_{i3}) & \cos(x_{i3}) & 0 \\  0 & 0 & 1
\end{bmatrix}J_{i}^{-\top}R_{i}u_{i}, 
\end{align}
where $u_{i}=(u_{i1}, u_{i2}, u_{i3})\in\mathbb{R}^{3}$ is the angular velocity of the wheels, $R_{i}=0.02$ is the wheel radius, and the matrix $J_{i}=\begin{bmatrix}\begin{smallmatrix}0 & \cos(\pi/6) & -\cos(\pi/6) \\ -1 & \sin(\pi/6) & \sin(\pi/6) \\ L_{i} & L_{i} & L_{i}\end{smallmatrix}\end{bmatrix}$ shows the geometric constraints with $L_{i}=0.2$ being the radius of the robot body. 

\begin{figure}[!t]
\begin{center}
\begin{picture}(70, 120)
\put(-70, -15){\resizebox{70mm}{50mm}{\includegraphics[width=2.5in]{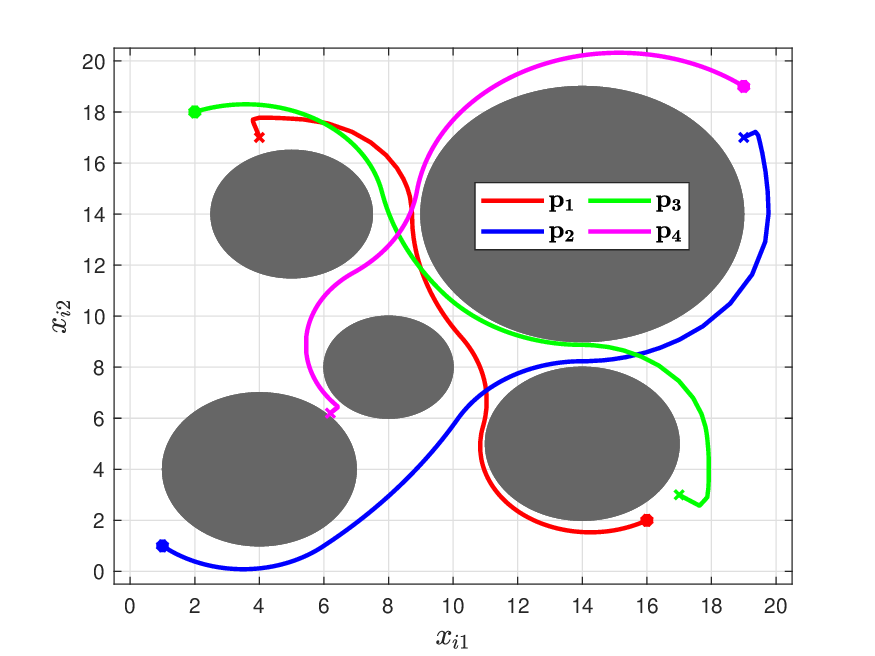}}}
\end{picture}
\end{center}
\caption{Illustration of the position trajectories of all robots via the optimization based control design. The dark grey regions are the obstacles, the initial positions are the dots, and the terminal positions are the crosses.}
\label{fig-1}
\end{figure}

All robots share the same position space where five obstacles appear; see Fig. \ref{fig-1}. Each robot aims to achieve its reach-avoid objective, which is to reach the terminal position while avoiding all obstacles. All terminal positions are given as $\mathfrak{q}_{i}\in\mathbb{R}^{2}$. For the reach objective, we define the MCLFs $V_{i}(x^{i}_{t})=V_{i1}(x_{i})+V_{i2}(x^{i}_{t})$ with $V_{i1}(x_{i})=(\mathfrak{p}_{i}-\mathfrak{q}_{i})^{\top}P_{i}(\mathfrak{p}_{i}-\mathfrak{q}_{i})$ and $V_{i2}(x^{i}_{t})=\sigma_{i}\int^{\Delta}_{0}(\mathfrak{p}^{i}_{t}(\theta)-\mathfrak{q}_{i})^{\top}Q_{i}(\mathfrak{p}^{i}_{t}(\theta)-\mathfrak{q}_{i})d\theta$, where $P_{i}, Q_{i}\in\mathbb{R}^{2\times2}$ are the positive definite matrices and $\sigma_{i}>0$. Here we choose $P_{i}=Q_{i}=I, \sigma_{1}=\sigma_{2}=0.1, \sigma_{3}=0.15$ and $\sigma_{3}=0.05$. We follow Theorem \ref{thm-1} to design the distributed stabilizing controller. In particular, we assume that the functionals $\rho_{i}, \gamma_{ij}$ in \eqref{eqn-5} are linear. That is, $\rho_{i}(s)=\bar{\rho}_{i}s$ with $\bar{\rho}_{i}=1$ and $\gamma_{ij}(s)=\bar{\gamma}_{ij}s$ with $\bar{\gamma}_{ij}=0.2$. Hence, item (iii) in Definition \ref{def-4} is satisfied. For each obstacle, its functional $h_{k}$ is defined as $h_{k}(\phi)=\mathsf{R}^{2}_{k}-(\mathfrak{p}-\mathbf{r}_{k})^{\top}(\mathfrak{p}-\mathbf{r}_{k})$, where $k=\{1, 2, 3, 4, 5\}$, $\phi\in\mathcal{PC}([-\Delta, 0], \mathbb{R}^{3})$ is the time-delay state with $\mathfrak{p}$ being the position, $\mathsf{R}_{k}>0$ is the radius and $\mathbf{r}_{k}\in\mathbb{R}^{2}$ is the center. 

\begin{figure}[!t]
\begin{center}
\begin{picture}(70, 120)
\put(-70, -15){\resizebox{70mm}{50mm}{\includegraphics[width=2.5in]{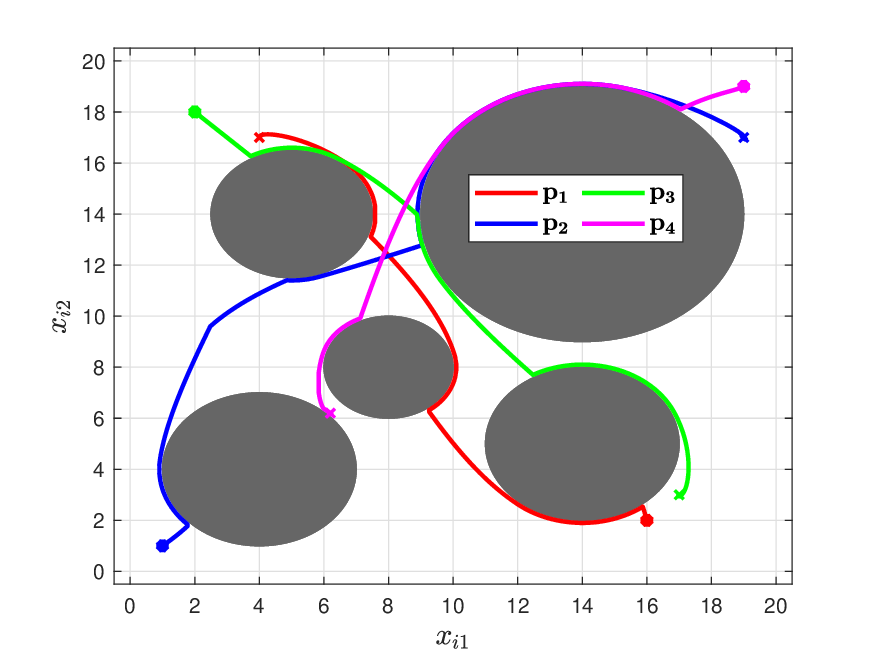}}}
\end{picture}
\end{center}
\caption{Illustration of the position trajectories of all robots via the sliding mode based control design. The dark grey regions are the obstacles, the initial positions are the dots, and the terminal positions are the crosses.}
\label{fig-2}
\end{figure}

Finally, we apply $h_{k}(\phi)$ to construct the MCBFs $B_{i}$ so that all the conditions in Definition \ref{def-6} are satisfied. Similar to the MCLFs, $\eta_{i}$ and $\chi_{ij}$ in \eqref{eqn-8} are assumed to be linear and set via the desired performance. To achieve all reach-avoid objectives, the first proposed control strategy is based on the optimization control as in \eqref{eqn-11}. Note that $u^{i}_{\mathsf{nom}}$ is from \eqref{eqn-6} via the MCLFs, and all CBF-based conditions are the constraints in \eqref{eqn-11}. By solving the optimization problem \eqref{eqn-11} using the \verb"fmincon" function in MATLAB, we derive the position trajectories as shown in Fig. \ref{fig-1}. On the other hand, we define the sliding mode functionals as $U_{i}(\phi)=V_{i}(\phi_{i})+\sum^{5}_{k=1}\varkappa_{ik}B_{k}(\phi_{i})$ with the weights $\varkappa_{ik}>0$, and then design the distributed controller as in \eqref{eqn-15}. In this way, the derived position trajectories of all robots are depicted in Fig. \ref{fig-2}. From Figs. \ref{fig-1}-\ref{fig-2}, we can see clearly the satisfaction of all reach-avoid objectives via the two proposed methods. Regarding these two methods, we remark that the MCBF-based conditions are listed item by item as the constraints in \eqref{eqn-11} while are combined together in the defined sliding mode functionals $U_{i}$. Hence, the position trajectories in Fig. \ref{fig-1} are smoother than these in Fig. \ref{fig-2}. However, the computation of \eqref{eqn-11} is much huger than the one via the sliding mode functionals, since the optimization problem needs to be solved in real time while the explicit form of the distributed controller is established in \eqref{eqn-15}. 

\section{Conclusion}
\label{sec-conclusion}

We proposed novel multiple control Lyapunov and barrier functionals to address safety and stabilization control problems of interconnected time-delay systems. The proposed multiple control functionals were further combined together using two different methods, and the distributed controllers were designed both implicitly and explicitly. The implicit controller was determined via a distributed convex quadratic program, while the explicit controller was based on the proposed sliding mode surface. Future work will be devoted to the construction of multiple control functionals and more general cases including motion control for multi-robot systems.


\end{document}